\documentclass[10pt, twocolumn]{article}

\usepackage{epsfig}
\usepackage{tabularx}
\usepackage{graphicx} 
\usepackage{color}
\usepackage{xspace}
\usepackage{thumbpdf}
\usepackage{listings}
\usepackage{verbatim}
\usepackage{hyperref}
\usepackage{colortbl}
\usepackage{scalerel,flushend}
\usepackage{authblk}
\usepackage{caption}
\usepackage{subcaption}

\begin{document}

\newcommand{\superscript}[1]{\ensuremath{^{\textrm{#1}}}}

\title{Miuz: measuring the impact of\\disconnecting a node}


\author[1]{Ivana Bachmann
\thanks{Email: \href{mailto:ivana@niclabs.cl}{ivana@niclabs.cl} To whom correspondence should be addressed.}}
\affil[1,4]{NICLabs, Computer Science Department (DCC), Universidad de Chile, Chile}
\author[2]{Patricio Reyes
\thanks{Email: \href{mailto:preyes@usc.es}{preyes@usc.es}}}
\affil[2]{Technological Institute for Industrial Mathematics (ITMATI), Santiago de Compostela, Spain}
\author[3]{Alonso Silva
\thanks{Email: \href{mailto:alonso.silva@nokia-bell-labs.com}{alonso.silva@nokia-bell-labs.com}}}
\affil[3]{Nokia Bell Labs, Centre de Villarceaux, Route de Villejust, 91620 Nozay, France}
\author[4]{Javier Bustos-Jim\'enez
\thanks{Email: \href{mailto:jbustos@niclabs.cl}{jbustos@niclabs.cl}}}

\date{}

\maketitle


\begin{abstract}
In this article we present \textit{Miuz}, a robustness index for complex networks.
\textit{Miuz} measures the impact of disconnecting a node from the network while comparing the sizes of the remaining connected components.
Strictly speaking, \textit{Miuz} for a node is defined as the inverse of the size of the largest connected component divided by the sum of the sizes of the remaining ones.

We tested our index in attack strategies where the nodes are disconnected in decreasing order of a specified metric. We considered \textit{Miuz} and other well-known centrality measures such as betweenness, degree, and harmonic centrality. All of these metrics were compared regarding the behavior of the robustness ($R$-index) during the attacks. In an attempt to simulate the internet backbone, the attacks were performed in complex networks with power-law degree distributions (scale-free networks).

Preliminary results show that attacks based on disconnecting a few number of nodes \textit{Miuz} are more dangerous (decreasing the robustness) than the same attacks based on other centrality measures. 
We believe that \textit{Miuz}, as well as other measures based on the size of the largest connected component, provides a good addition to other robustness metrics for complex networks.
\end{abstract}

\section{Introduction}
Networks are present in human life in multiple forms from social networks to communication networks (such as the Internet), and they have been widely studied as complex networks of nodes and relationships, describing their structure,  relations,  etc.  Nowadays, the studies go deep into network knowledge, and using standard network metrics such as degree distribution and diameter one can determine how robust and/or resilient a network is.

Even though both terms (robustness and resilience) have been used with the same meaning, we will consider \textit{robustness} as the network inner capacity to resist failures, and \textit{resilience} as the ability of a network to resist and recover after such failures. However, both terms have in common the methodology for testing robustness and/or resilience. Usually, they consist on planned attacks against nodes failures or disconnections, from a random set of failures to more elaborated strategies using well known network metrics.  

We consider that an ``adversary'' should plan a greedy strategy aiming to maximize damage with the minimum number of strikes. Under this new philosophy of ``adversary'', we first present a new network impact metric called \textit{Miuz}, which is the inverse of the size of the largest connected component, divided by the sum of the sizes of the remaining ones. Then, we plan a greedy strategy based on recalculating the network \textit{Miuz-ness} after each node disconnection. We discuss the performance of attacks based on \textit{Miuz} and other centrality measures\cite{bersano2012metrics} (in particular, betweenness, degree, harmonic), compared by the robustness index ($R$-index). Our main conclusion is that the first strikes of a \textit{Miuz} strategy causes more ``\textit{damage}'' than the classical betweenness, degree, and harmonic metrics.

The article is organized as follows, next section presents related work, followed by the definition of \textit{Miuz}, its attacking strategy, and its simulation results in section \ref{miuz}. Discussions about the main results and conclusions are presented in sections \ref{discussion} and \ref{conclusions} respectively.

\section{Related Work}
\label{related}
Over the last decade, there has been a huge interest in the analysis
of complex networks and their connectivity properties~\cite{Albert2000}. During the last years, networks
and in particular social networks have gained significant popularity. An
in-depth understanding of the graph structure is key to convert data
into information. To do so, complex networks tools have emerged~\cite{Albert2002} to classify networks~\cite{Watts1998},
detect communities~\cite{Leskovec2008}, determine
important features and measure them~\cite{bersano2012metrics}.

The idea of planning a ``network attack'' using centrality measures has been catching the attention of researchers and practitioners nowadays. For instance, Sterbenz et al.~\cite{sterbenz2011modelling} used bet\-ween\-ness-centrality (\textit{bcen}) for planning a network attack, calculating the \textit{bcen} value for all nodes, ordering nodes from higher to lower \textit{bcen}, and then attacking (discarding) those nodes in that order. They have shown that disconnecting only two of the top \textit{bcen}-ranked nodes %
, their packet-delivery ratio is reduced to $60\%$, which corresponds to~$20\%$ more damage than other attacks such as random links or nodes disconnections, tracked by link-centrality and by node degrees.

A similar approach and results were presented by {\c{C}}etin\-kaya et al.~\cite{ccetinkaya2013modelling}. They show that after disconnecting only $10$ nodes 
the packet-delivery ratio is reduced to $0\%$. Another approach, presented as an improved network attack \cite{rak2010survivability, sydney2010characterising}, is to recalculate the betweenness-centrality after the removal of each node \cite{holme2002attack,molisz2006end}. They show a similar impact of non-recalculating strategies but discarding sometimes only half of the equivalent nodes. 

Concerning centrality measures, betweenness centrality deserves special attention.
Betweenness has been studied as a resilience metric for the routing layer~\cite{smith2011network} and also as a robustness metric for complex networks \cite{iyer2013attack} and for internet autonomous systems networks~\cite{mahadevan2006internet} among others.

\section{Miuz Attacking Strategy}
\label{miuz}

\begin{figure*}[t!]
	\centering
	\begin{subfigure}[b]{0.3\textwidth}
		\includegraphics[width=\textwidth, natwidth=300,natheight=300]{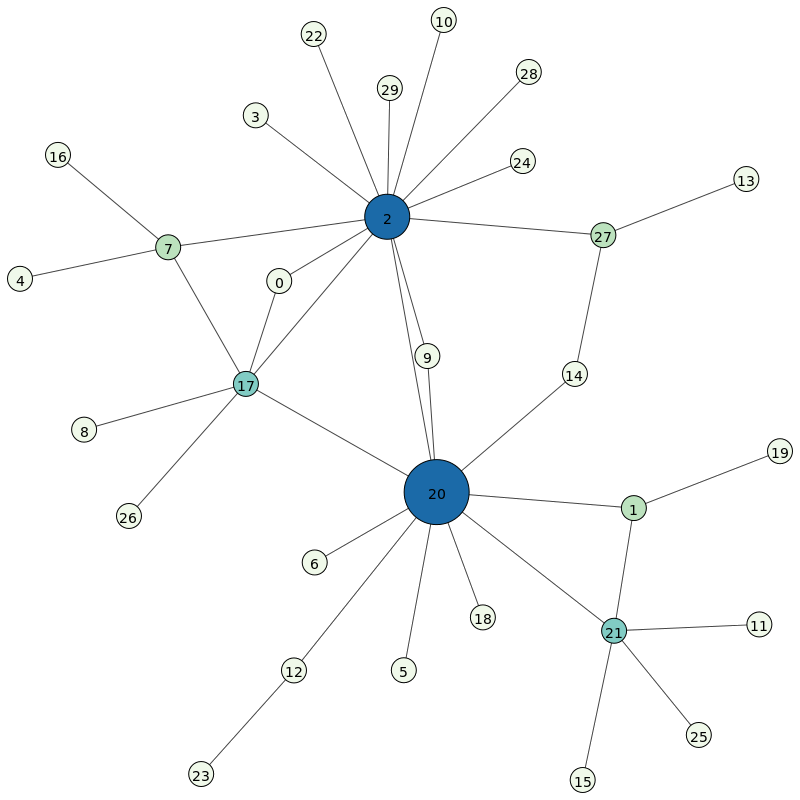}
		\caption{Original network A, color coded by degree  centrality}
	\end{subfigure}
	~
	\begin{subfigure}[b]{0.3\textwidth}
		\includegraphics[width=\textwidth, natwidth=300,natheight=300]{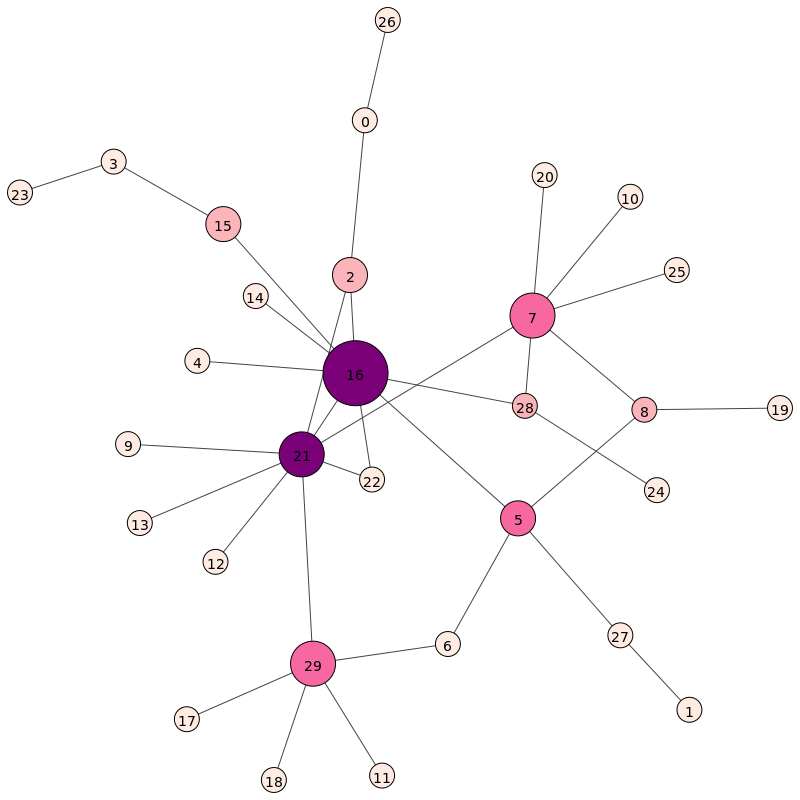}
		\caption{Original network B, color coded by betweenness centrality}
	\end{subfigure}
	~
	\begin{subfigure}[b]{0.3\textwidth}
		\includegraphics[width=\textwidth, natwidth=300,natheight=300]{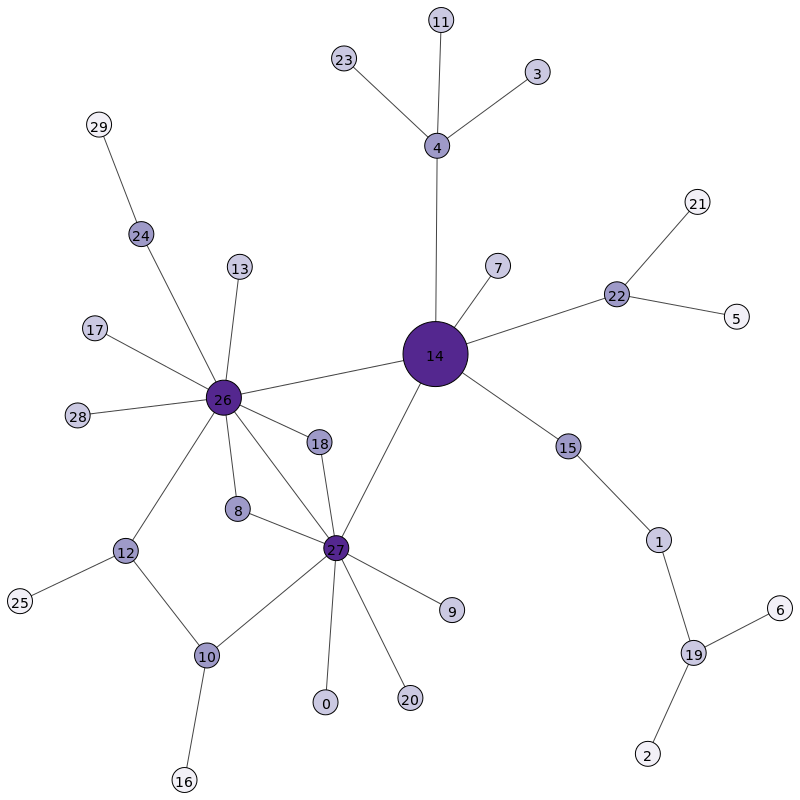}
		\caption{Original network C, color coded  by harmonic  centrality}
	\end{subfigure}
	~
	\begin{subfigure}[b]{0.3\textwidth}
		\includegraphics[width=\textwidth, natwidth=300,natheight=300]{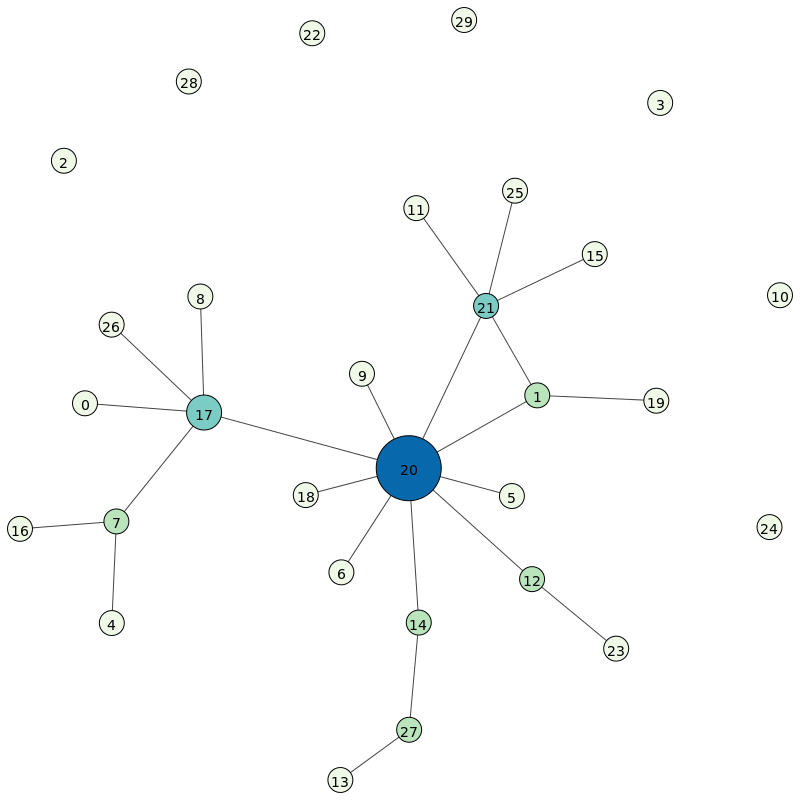}
		\caption{Network A, after removal of node with highest degree centrality}
	\end{subfigure}
	~
	\begin{subfigure}[b]{0.3\textwidth}
		\includegraphics[width=\textwidth, natwidth=300,natheight=300]{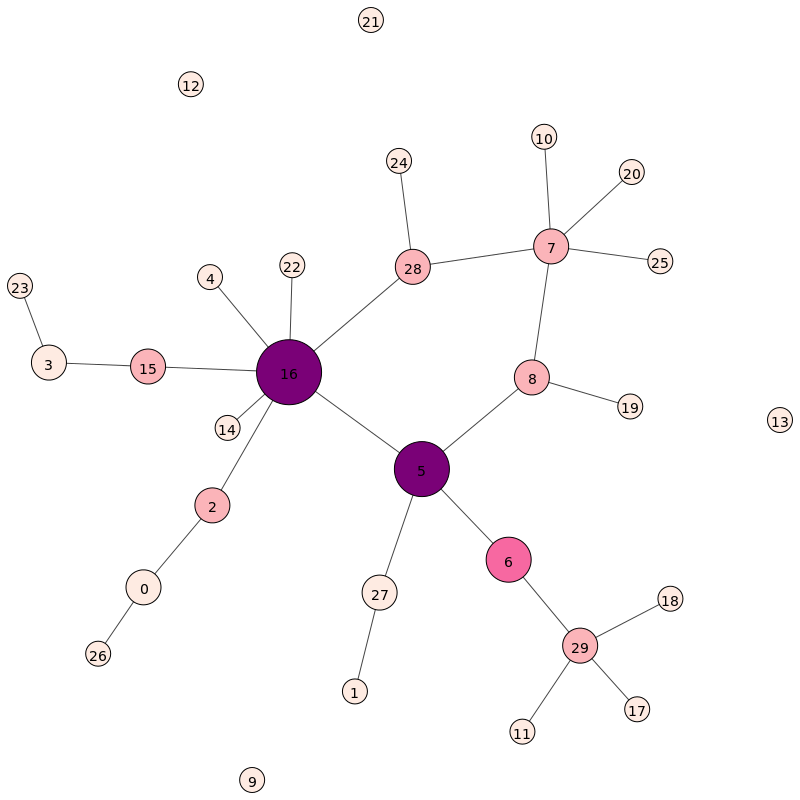}
		\caption{Network B, after removal of node with highest betweenness centrality}
	\end{subfigure}
	~
	\begin{subfigure}[b]{0.3\textwidth}
		\includegraphics[width=\textwidth, natwidth=300,natheight=300]{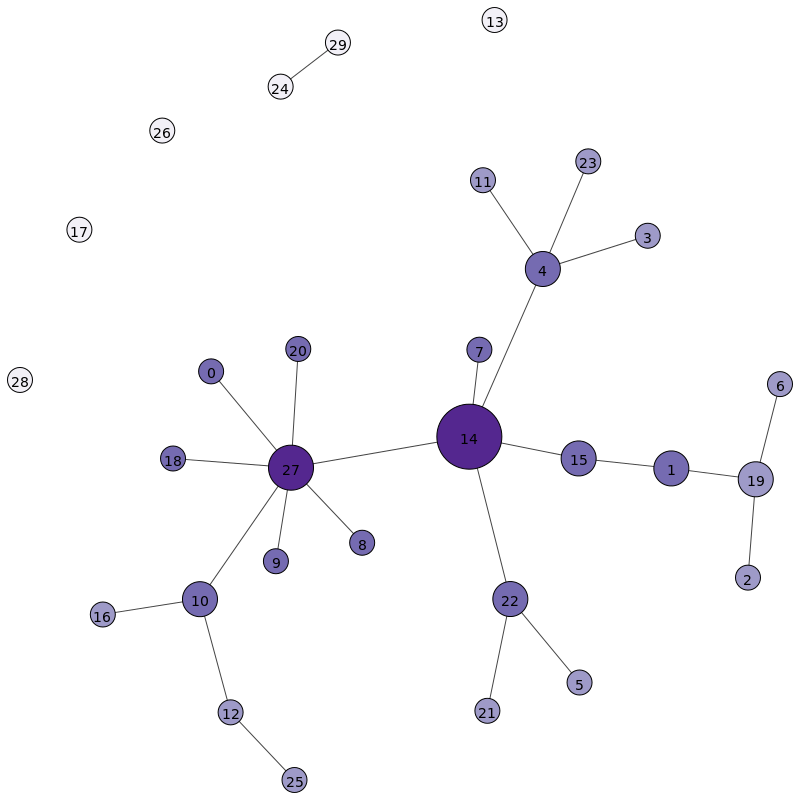}
		\caption{Network C, after removal of node with highest harmonic centrality}
	\end{subfigure}
	~
	\begin{subfigure}[b]{0.3\textwidth}
		\includegraphics[width=\textwidth, natwidth=300,natheight=300]{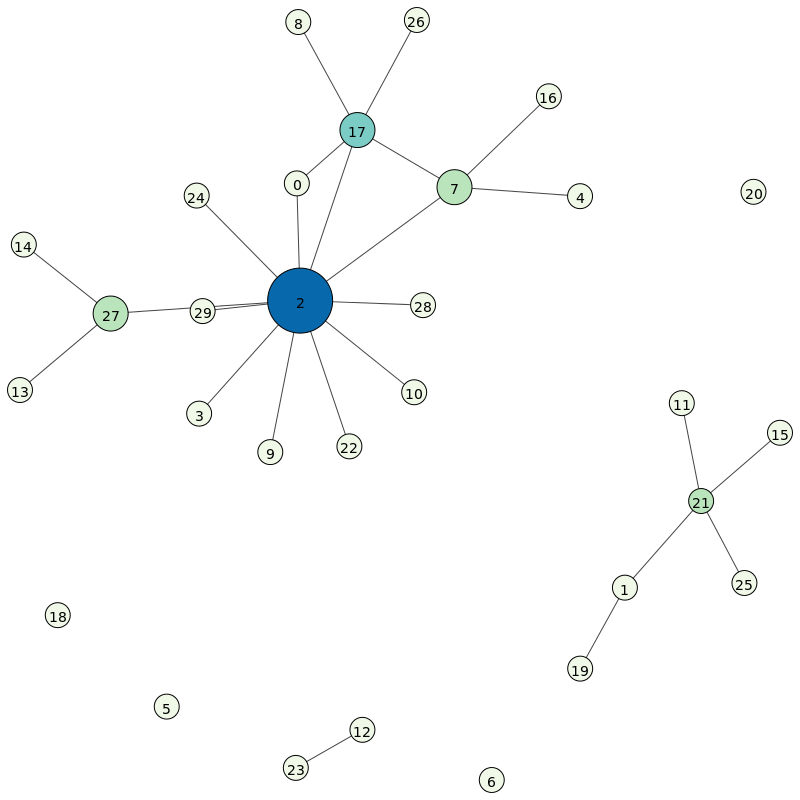}
		\caption{Network A, after removal of node with highest \textit{Miuz}}
	\end{subfigure}
	~
	\begin{subfigure}[b]{0.3\textwidth}
		\includegraphics[width=\textwidth, natwidth=300,natheight=300]{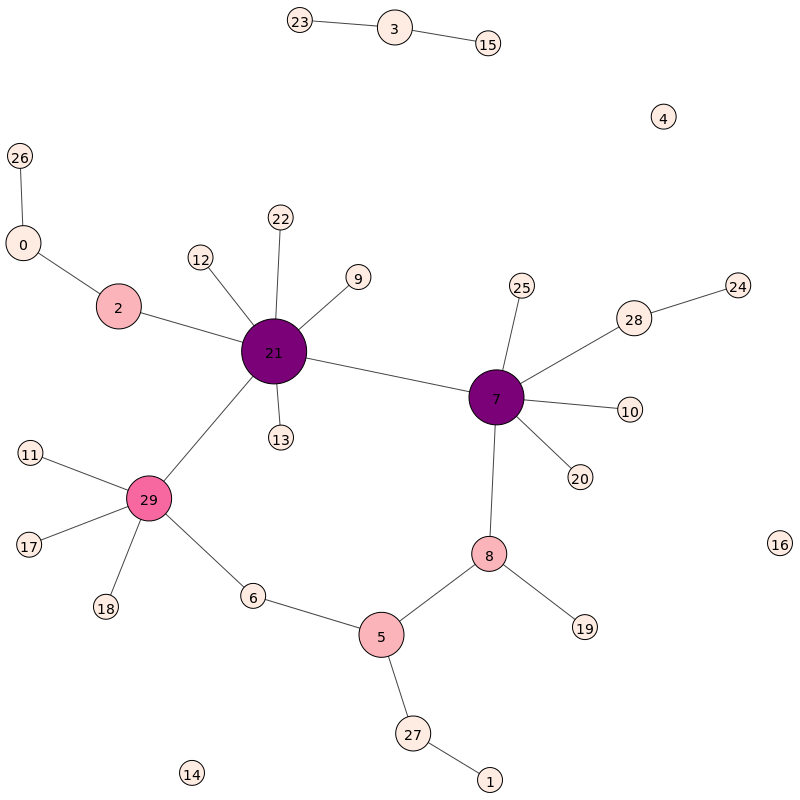}
		\caption{Network B, after removal of node with highest  \textit{Miuz}}
	\end{subfigure}
	~
	\begin{subfigure}[b]{0.3\textwidth}
		\includegraphics[width=\textwidth, natwidth=300,natheight=300]{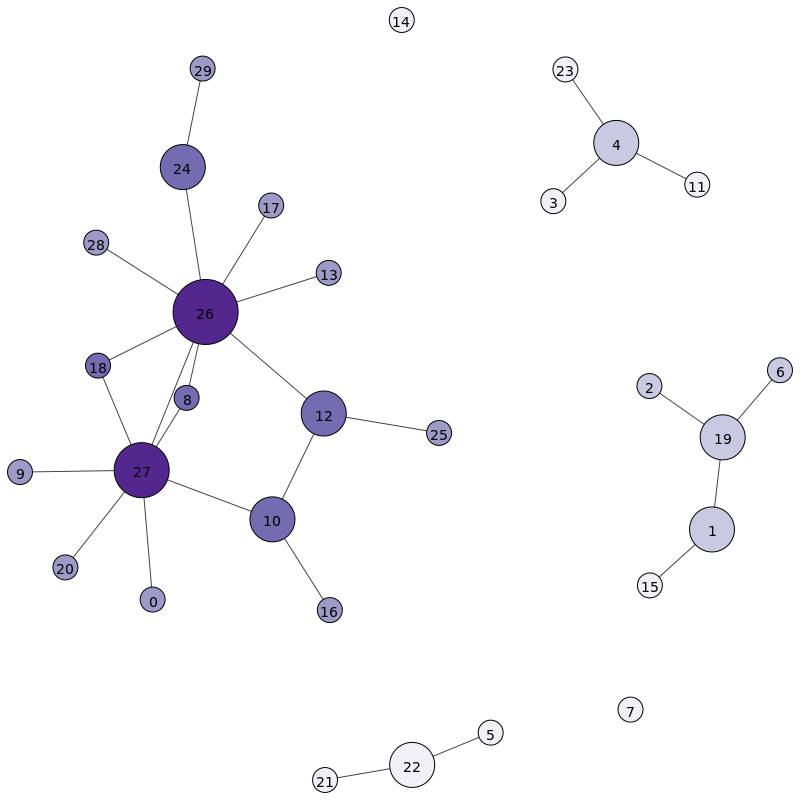}
		\caption{Network C, after removal of node with highest  \textit{Miuz}}
	\end{subfigure}
 	\caption{Effects of attacking strategies. The size shows the Miuz-ness and the color darkness reflects the centrality metric. Metrics were recalculated after node removal.}
	\label{fig:figure3}
\end{figure*}

Given a network $\mathcal{N}$ of size $N$,  we denote by $\mathcal{C}(\mathcal{N} \setminus n)$ the set of connected components in $\mathcal{N}$ after disconnecting node~$n$. The \textit{Miuz} index for a node $n$ in $\mathcal{N}$ is defined as follows:
\begin{equation}
{\scriptsize
\textit{Miuz}_{\mathcal{N}}(n) = 
  \left\{
    \begin{array}{l}
      \frac{\sum_{\textit{c} \in \mathcal{C} (\mathcal{N}\setminus n)}{\|\textit{c}}\|}{max_{\textit{c} \in \mathcal{C} (\mathcal{N}\setminus n)}\|c\|} - 1, \textit{if }  \|\mathcal{C} (\mathcal{N}\setminus n)\| \neq  \|\mathcal{C} (\mathcal{N})\| + 1\\
       \\
      0 \hfill \textit{otherwise}
    \end{array}
  \right.
}
\end{equation}

\noindent where $\|c\|$, with $\textit{c} \in \mathcal{C} (\mathcal{N}\setminus n)$, is the size of the connected component $c$ of the network $\mathcal{N}$ after disconnecting node $n$ (if connected, i.e. if there is an edge between node $n$ and another node of the network). Notice that $\textit{Miuz}_\mathcal{N}$(n) reflects the partition of a network in several sub-networks after the disconnection of node $n$ and how these sub-networks remain interconnected. Strictly speaking, it compares in size the core network (the largest connected component) with the other remaining sub-networks. $\textit{Miuz}_\mathcal{N}$ takes values between $0$ (the whole network remains connected) to $N-1$ (the whole network is disconnected). 
Figure~\ref{fig:figure3} shows examples of three networks after disconnecting the node with either highest \textit{Miuz}, degree, betweenness or harmonic centrality. 

\subsection{Network Attack Plan}

If we plan a network attack by disconnecting nodes with a given strategy, it is widely accepted the use of centrality measures (such as degree, betweenness, harmonic) because they reflect the importance of nodes in the network \cite{iyer2013attack}. These attack strategies are compared by means of the \textit{Unique Robustness Measure} ($R$-index)~\cite{schneider2011mitigation}, defined as: 
\begin{equation}
R = \frac{1}{N}\sum_{Q=1}^{N} {s(Q)},
\end{equation}
where $N$ is the number of nodes in the network and $s(Q)$ is the fraction of nodes in the largest connected component after removing $Q$ nodes using a given strategy.  Therefore, the higher the $R$-index, the better in terms of robustness. Our strategy is:\\

\textit{In each step, (re)compute the \textit{Miuz} index for all the nodes and disconnect the node with the highest \textit{Miuz} value.} \\

Degree centrality is a simple, easy-to-compute local (myopic) metric. Despite its simplicity, it is the preferred metric for certain literature~\cite{iyer2013attack} under this kind of attacks. Basically, it ranks nodes in terms of their degree (number of edges or connections to other nodes). Betweenness centrality ranks nodes by the number of shortest paths passing throw them. Unlike degree, betweenness centrality is not a local (myopic) metric. It determines the importance of a node by looking at the paths between all of the pairs of remaining nodes. The harmonic centrality is a metric related to the distances to the remaining nodes. It is the sum of the reciprocal of distances to other nodes (with the convention of $1/\infty = 0$, used when two nodes are not connected by any path). This metric was widely studied in~\cite{Boldi2014}, mainly because of its good mathematical properties.

For a better understanding of network attacks and strategies, see  \cite{holme2002attack,molisz2006end,rak2010survivability, sydney2010characterising}. 

\subsection{Simulations}

In an attempt to study the behavior of the internet backbone, we test our strategy in simulated scale-free networks (it means, with power-law degree distributions, that is $\textit {dist(deg)} \propto x^{-\alpha}$) with exponents $\alpha \in \{2.10,2.15,2.20,2.25,2.30\}$. For each $\alpha$, we simulate $50$ networks of size $1000$ (i.e., with $1000$ nodes).
Then, we tested and compared strategies ranked with a multitude of centrality measures~\cite{bersano2012metrics} in terms of the $R$-index. However, apart from \textit{Miuz}, for this article we selected and reported the results for degree, betweenness and harmonic centrality.

Instead of just compare the robustness, after the removal of all of the nodes, we studied the behavior of the attacks after the only a few strikes. To do so, we define a variant of the $R$-index which takes into account only the first $n$ strikes of an attack. Thus, for a simultaneous attack (where the nodes are ranked by a metric only once at the beginning), the $R_a$-index is defined as:
\begin{equation}
R_a = \frac{1}{a}\sum_{Q=1}^{a} {s(Q)}.
\end{equation}
For a sequential attack, the order of node disconnection is recomputed after each disconnection. Similar to the $R$-index, notice that the lower the $R_a$-index, the more effective the attack is, since that gives us a higher reduction of robustness. 

Results are shown in Table~\ref{tab:table2}. We tested sequential attacks: At each strike, the next node to disconnect was the one with the highest metric (whether it be \textit{Miuz}, Degree, Betweenness or Harmonic centrality) in the current network. The table shows the behavior of the $R_a$-index as well as the $R$-index in 50 scale-free networks (generated as before) with exponent 2.1, 2.2 and 2.3. For each group of networks, we compute the $R_a$-index for $a \in \{5,10,20,30\}$, it means, the variant of $R$-index after the first $5,10,20$ and $30$ node disconnections. \textit{Miuz} proves to be very effective in attacks with only a few disconnections. Moreover, \textit{Miuz} shows that the effectiveness persists over the number of strikes in scale-free networks with lower exponent.

It is interesting to note that, no matter the metric used, the damage decreases in the long term with the number of strikes. Indeed, if we consider a complete attack, disconnecting all the nodes of the network, the $R$-index shows that an attack based on Harmonic centrality performs better. A comparison of the $R$-index is shown in Table~\ref{tab:table1}. It is important to notice that $R$-index is a metric for a general view of robustness, and it gives no information of how fast the network is disconnected (further details in Section \ref{R}).

\begin{table*}[ht]
\centering
\caption{Robustness against sequential attacks based on different metrics. Table shows mean $R$-index and mean $R_a$-index for scale-free networks with exponent $2.1$, $2.2$ and $2.3$. $R_a$-index corresponds to the equivalent to $R$-index, but only disconnecting $a$ nodes. Attacks based on harmonic centrality are more harmful in the long term. However attacks based on \textit{Miuz} turns out to be more dangerous than the others with only a few strikes.}
\label{tab:table2}
\begin{tabular}{|l|r|r|r|r|}
\hline
metric   & \multicolumn{1}{l|}{Miuz} & \multicolumn{1}{l|}{Degree} & \multicolumn{1}{l|}{Betweenness} & \multicolumn{1}{l|}{Harmonic} \\ \hline
\multicolumn{5}{|c|}{scale-free network with exponent 2.1}                                                                                                 \\ \hline
$R_5$-index    & {\bf 0.6976}              & 0.6987                      & 0.6994                           & 0.6999                        \\ \hline
$R_{10}$-index & {\bf 0.6068}              & 0.6092                      & 0.6108                           & 0.6118                        \\ \hline
$R_{20}$-index & {\bf 0.4929}              & 0.499                       & 0.5032                           & 0.5059                        \\ \hline
$R_{30}$-index & {\bf 0.4051}              & 0.4135                      & 0.4092                           & 0.4166                        \\ \hline
$R$-index      & 0.016                     & 0.0155                      & 0.015                            & {\bf 0.0146}                  \\ \hline
\multicolumn{5}{|c|}{scale-free network with exponent 2.2}                                                                                                 \\ \hline
$R_5$-index    & {\bf 0.6867}              & 0.6886                      & 0.6897                           & 0.6903                        \\ \hline
$R_{10}$-index & {\bf 0.599}               & 0.6039                      & 0.6078                           & 0.6092                        \\ \hline
$R_{20}$-index & {\bf 0.4721}              & 0.4859                      & 0.4841                           & 0.4911                        \\ \hline
$R_{30}$-index & 0.3602                    & 0.3812                      & {\bf 0.3504}                     & 0.3619                        \\ \hline
$R$-index      & 0.0142                    & 0.0137                      & 0.0129                           & {\bf 0.0125}                  \\ \hline
\multicolumn{5}{|c|}{scale-free network with exponent 2.3}                                                                                                 \\ \hline
$R_5$-index    & {\bf 0.6601}              & 0.677                       & 0.6708                           & 0.6734                        \\ \hline
$R_{10}$-index & {\bf 0.5321}              & 0.5585                      & 0.5449                           & 0.5541                        \\ \hline
$R_{20}$-index & 0.3551                    & 0.3926                      & {\bf 0.3405}                     & 0.3567                        \\ \hline
$R_{30}$-index   & 0.2586                    & 0.2862                      & {\bf 0.2363}                     & 0.2487                        \\ \hline
$R$-index      & 0.0123                    & 0.1063                      & 0.0094                           & {\bf 0.0091}                  \\ \hline
\end{tabular}
\end{table*}

\begin{table*}[ht]
\begin{center}
\begin{tabular}{|c||c|c|c|c|}\hline
$\alpha$   & Miuz & Degree & Betweenness & Harmonic \\\hline
2.10  & 0.01603 & 0.01549 & 0.01496 & \textbf{0.01460}  \\\hline
2.15  & 0.01575 & 0.01525 & 0.01430 & \textbf{0.01400}  \\\hline
2.20 & 0.01421& 0.01372 & 0.01287 & \textbf{0.01249} \\\hline
2.25 & 0.01384& 0.01299 & 0.01874 & \textbf{0.01150} \\\hline
2.30  & 0.01232& 0.010631 & 0.00942 &\textbf{0.009094}  \\\hline
\end{tabular}
\end{center}
\caption{Mean values of $R$-index for different disconnection strategies. The harmonic centrality strategy gets the lowest $R$-index values (in bold) for all of the tested exponents. (The higher the $R$-index, the better in terms of robustness).}
\label{tab:table1}
\end{table*}%

\section{Discussion}
\label{discussion}

\begin{figure*}[t!]
	\centering
	\begin{subfigure}[b]{.5\textwidth}
		\includegraphics[width=\textwidth, natwidth=300,natheight=300]{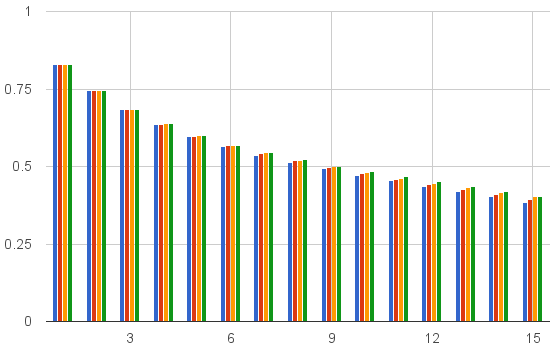}
		\caption{$\alpha=2.10$}
	\end{subfigure}
	~
	\begin{subfigure}[b]{.5\textwidth}
		\includegraphics[width=\textwidth, natwidth=300,natheight=300]{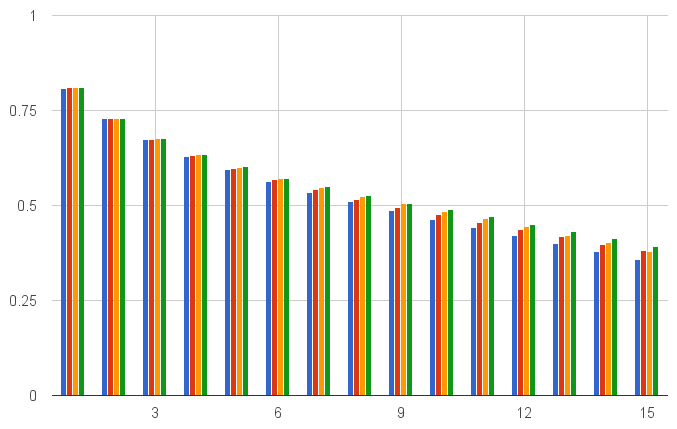}
		\caption{$\alpha=2.20$}
	\end{subfigure}
	~
	\begin{subfigure}[b]{.5\textwidth}
		\includegraphics[width=\textwidth, natwidth=300,natheight=300]{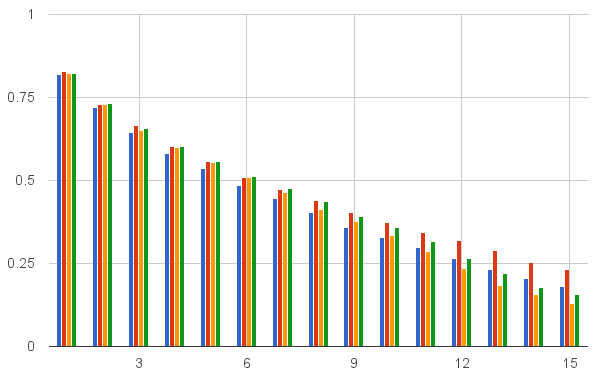}
		\caption{$\alpha=2.30$}
	\end{subfigure}
 	\caption{Effects of sequentially disconnecting nodes on power-law networks, plotting the fractional size of largest component after disconnecting $x$ nodes and using as a selection strategy: Miuz (blue), Degree (red), Betweenness (orange), and Harmonic (green).}
	\label{fig:figure1}
\end{figure*}

We start analyzing what would be the worst ``attack''. In the worst case, a ``\textit{malicious adversary}'' will try to perform the maximum damage with the minimum number of strikes, that is, with the minimum number of node disconnections (made by the attacker).
In terms of decreasing the size of the largest connected component, \textit{Miuz} is the best attack strategy during the first strikes. It achieves the disconnection of more than half of the network only after nine strikes for $\alpha=2.10$ and after five strikes for $\alpha=2.30$ (see Fig. \ref{fig:figure1}).

\begin{figure*}[t!]
	\centering
	\begin{subfigure}[b]{.5\textwidth}
		\includegraphics[width=\textwidth, natwidth=300,natheight=300]{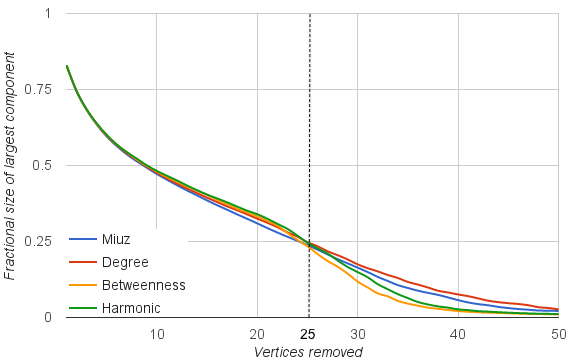}
		\caption{$\alpha=2.1$}
	\end{subfigure}
	~
	\begin{subfigure}[b]{.5\textwidth}
		\includegraphics[width=\textwidth, natwidth=300,natheight=300]{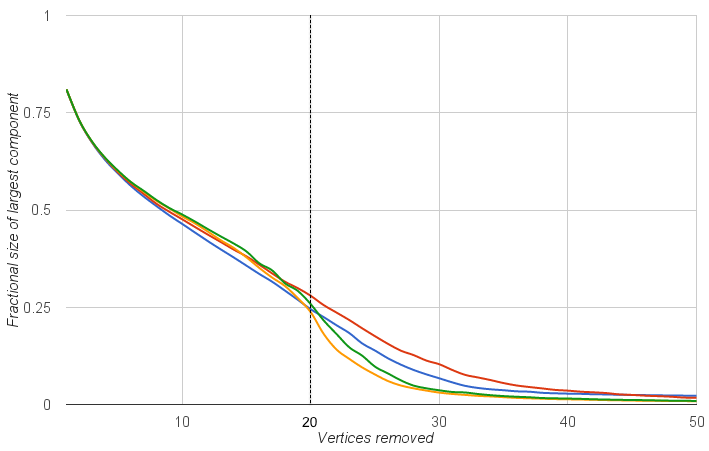}
		\caption{$\alpha=2.2$}
	\end{subfigure}
	~
	\begin{subfigure}[b]{.5\textwidth}
		\includegraphics[width=\textwidth, natwidth=300,natheight=300]{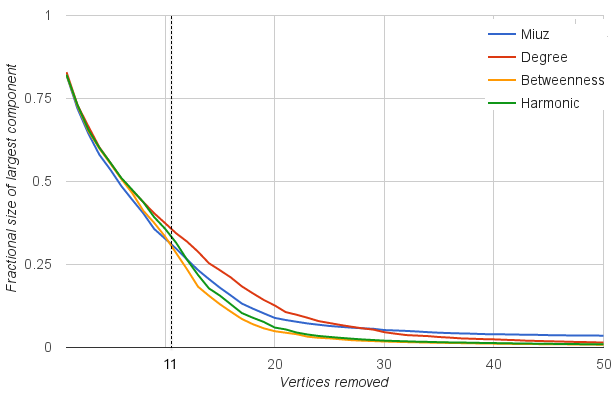}
		\caption{$\alpha=2.3$}
	\end{subfigure}
 	\caption{Size of the largest fractional connected component and number disconnected nodes. Dashed line shows the breaking point when Miuz is no longer the best attacking strategy for $\alpha=2.1-2.3$.}
	\label{fig:figureX}
\end{figure*}

It is important to notice that there is a breaking point where \textit{Miuz} is no longer the best attacking strategy (Fig. \ref{fig:figureX}). This breaking point occurs with less strikes as the exponent of the scale-free network increases. Nevertheless, as shown in Fig. \ref{fig:figureX}, this breaking point appears after the largest connected component is less than $1/3$ of the original network.

\subsection {Is the $R$-index a good metric for robustness?}
\label{R}

From the above discussion it is valid to question whether the $R$-index is a good measure of robustness. Maybe it is important to revisit the definition of network robustness: 

\textit{``A measure of the decrease of network functionality under a sinister attack''}\cite{holme2002attack}

Therefore, the $R$-index can be considered as a good robustness metric even though it does not reflect how fast the network is disconnected (as it is shown in Fig. \ref{fig:figure2}). 

We suggest that weighted versions of the $R$-index could be a good compromise. This is the case of the $R_a$-index proposed in Section~\ref{miuz} which takes into account the removal of only a percentage of the best ranked nodes (or until the network reaches its percolation threshold \cite{holme2002attack}) could be a more useful metric for robustness and/or resilience in worst-case scenarios.%
 

\subsection {Revisiting Miuz}

Concerning the attacking strategies, it is important to notice that \textit{Miuz} strategy was designed for disconnecting the network from the first strikes, aiming to get connected components with similar sizes. This is the main reason of its better performance than centrality metrics in a worst-attack scenario. For instance, in Fig. \ref{fig:figure3} we present  examples where the \textit{Miuz} strategy performs better attacks than the ones with other centrality metrics. In subfigure a) Degree-based attack will select node $2$ (up) while \textit{Miuz} selects node $20$ (down). In b) Betweenness-based attack will select node $21$ while \textit{Miuz} selects node $18$. In c) Harmonic-based attack will select node $25$ (left) while \textit{Miuz} selects node $14$ (right).  

Another interesting property of $\textit{Miuz}$ occurs when $\textrm{max}_n (\textit{Miuz}_n(\mathcal{N})) = 0$. In other words, when \textit{Miuz} is 0 for any node. In that case, the full network will remain connected no matter which node is disconnected. Therefore, we suggest a resilience metric as the number (or percentage) of disconnected nodes until $\textrm{max} (\textit{Miuz}_n(\mathcal{N})) > 0$, a pre-defined threshold, or the percolation threshold \cite{holme2002attack}.


\section{Conclusions and Future Work}
\label{conclusions}

In this article we have presented \textit{Miuz}, a robustness index for complex networks defined as the inverse of the size of the remaining largest connected component divided by the sum of the sizes of the remaining connected components.

We tested our index as a measure to quantify the impact of node removal in terms of the network robustness metric $R$-index. We compared \textit{Miuz} with other attacks based on well-known centrality measures (betweenness centrality, degree and harmonic) for node removal selection. The attack strategy used was sequential targeted attack, where every index is recalculated after each removal, and the highest one is selected for the next extraction. 

Preliminary results show that \textit{Miuz} performs better compared to attacks with classical centrality measures in terms of decreasing the network robustness in the worst-attack scenario. Compared to other centrality measures, strategies based on \textit{Miuz} are more dangerous (decreasing the robustness) in the first strikes of the attacks. We suggest that \textit{Miuz}, as well as other measures based on the size of the largest connected component, provides a good addition to other robustness metrics for complex networks.

As future work, 
we can study improving through new connections an already existing network to make it more robust to attacks, and we can study networks in which the topological space is correlated with the cost of nodes disconnections, nodes which are nearby have a lower cost to be disconnected compared with nodes which are far apart.

\bibliographystyle{hieeetr} 
\bibliography{mybib}
\label{last-page}

\end{document}